\documentclass{article}%
\usepackage{amsfonts}
\usepackage{amsmath}
\usepackage{amssymb}
\usepackage{graphicx}%
\begin{document}

\title{The Schwarzschild-de Sitter solution in five-dimensional general relativity
briefly revisited }
\author{J. B. Fonseca-Neto and C. Romero\\Departamento de F\'{\i}sica, Universidade Federal da Para\'{\i}ba,\\C.Postal 5008, 58051-970 Jo\~{a}o Pessoa, Pb, Brazil\\E-mail: cromero@fisica.ufpb.br }
\maketitle

\begin{abstract}
We briefly revisit the Schwarzschild-de Sitter solution in the context of
five-dimensional general relativity. We obtain a class of five-dimensional
solutions of Einstein vacuum field equations into which the four-dimensional
Schwarzschild-de Sitter space can be locally and isometrically embedded. We
show that this class of solutions is well-behaved in the limit of
$\Lambda\rightarrow0$. Applying the same procedure to the de Sitter
cosmological model in five dimensions we obtain a class of embedding spaces
which are similarly well-behaved in this limit. These examples demonstrate
that the presence of a non-zero cosmological constant does not in general
impose a rigid relation between the $(3+1)$ and $(4+1)$-dimensional
spacetimes, with degenerate limiting behaviour.

\end{abstract}

\section{Introduction}

The accumulation of recent observations has led to the so-called `standard'
cosmological model according to which the universe is presently dominated by a
`dark energy', a favourite candidate for which is the cosmological constant.
This has provided an urgent motivation to understand the nature of such a
constant. In addition to having important consequences for the cosmological
dynamics, such a constant would also in principle have consequences for local
dynamics. The well-known $(3+1)$-dimensional Schwarzschild-de Sitter
spacetime, discovered by de Sitter in 1917 in the cosmological context,
describes the gravitational field of a spherically symmetric body in presence
of a cosmological constant $\Lambda$. However, the influence of a small
non-zero cosmological constant on the usual solar system tests based on the
standard Schwarzschild metric, though in principle detectable, has been shown
to be very small \cite{Kagramanova}. Over the recent years increasing emphasis
has been placed on the possibility that the universe may in fact be
higher-dimensional than the usual 4-dimensional (4-D) Einsteinian one. In the
eighties, following the revival of Kaluza-Klein theory, some authors began
investigating the one-body problem in the context of five-dimensional $(4+1)$
general relativity \cite{Gross}. An interesting extension of the
$(3+1)$-dimensional Schwarzschild-de Sitter solution to five dimensions was
then carried out by Wesson and coworkers \cite{Wesson1}. In their analysis of
the one-body problem they addressed the question of how an extra dimension
would affect the classical tests of general relativity such as the perihelion
advance of Mercury, the gravitational redshift effect and the deflection of
light rays by the Sun. By comparing the geodesic motions in the corresponding
$(4+1)$ and $(3+1)$-dimensional spacetimes they concluded that as far as the
classical tests of general relativity are concerned there is no way to
distinguish between the 5-D and 4-D descriptions \cite{Wesson2}. This
conclusion, however, depends on the assumption that the metric of the 5-D
embedding space has a certain (canonical) specific form, which allows the 5-D
and 4-D geodesics to coincide. We shall return to this question below.

The solution due to Wesson and co-workers mentioned above corresponds to a
class of 5-D Ricci-flat spaces $\{M_{\Lambda}^{5}\}$, where each $M_{\Lambda
}^{5}$ is foliated by a family of four-dimensional hypersurfaces
$\{\Sigma_{\Lambda}^{4}\}$ which are locally and isometrically embedded in
$M_{\Lambda}^{5}$. It has been shown that $(3+1)$-dimensional Schwarzschild-de
Sitter spacetime $M_{\Lambda}^{4}$ belongs to the class $\{\Sigma_{\Lambda
}^{4}\}$ \cite{Kasner}.\ This is in accordance with the Campbell-Magaard
theorem which asserts that any analytic n-dimensional Riemannian
(pseudo-Riemannian) manifold can be locally embedded in an (n+1)-dimensional
Ricci-flat Riemannian (pseudo-Riemannian) manifold
\cite{Campbell,romero-etal,Seahra,Dahia}.

As was pointed out in Ref. \cite{Wesson2} the metric $g_{ab}(\Lambda)$\ of the
embedding space $M_{\Lambda}^{5}$ is degenerate in the limit $\Lambda
\rightarrow0$. Taken at face value, and assuming that this degeneracy is
generic, this might be taken to imply that the presence of non-zero
cosmological constant in $(3+1)$-dimensional spacetime may be tightly related
to its presence in the embedding $(4+1)$-dimensional spacetime. This raises
the interesting question of whether there exist different classes of
$(4+1)$-dimensional Ricci-flat spaces $\{\overline{M}_{\Lambda}^{5}\}$ with
metrics $\overline{g}_{ab}(\Lambda)$\ in which one can locally and
isometrically embed the Schwarzschild-de Sitter spacetime $M_{\Lambda}^{4}$,
with $\overline{g}_{ab}(\Lambda)$ possessing a well-defined nonvanishing limit
as $\Lambda\rightarrow0$. The main purpose of this paper is to give an
affirmative answer to this question and to exhibit the desired embedding. The
paper is organized as follows. In Section 2 we briefly review the one-body
solution in $(4+1)$-dimensions found by Wesson and collaborators. We then
proceed in Section 3 to give a new class of Ricci-flat spacetimes which
includes the latter solution as a particular case while remaining well-behaved
in the limit $\Lambda\rightarrow0$. Finally, in Section 4, we discuss the
embedding of the cosmological de Sitter model into a family of
five-dimensional flat spaces that are also well-behaved with respect to the
limit $\Lambda\rightarrow0$.

\section{The extension of Schwarzschild-de Sitter spacetime to five
dimensions}

The Schwarzschild-de Sitter spacetime corresponds to a 4-D manifold with the
spherically-symmetric line element
\begin{equation}
ds^{2}=\left(  1-\frac{2m}{r}-\frac{\Lambda r^{2}}{3}\right)  dt^{2}-\left(
1-\frac{2m}{r}-\frac{\Lambda r^{2}}{3}\right)  ^{-1}dr^{2}-r^{2}(d\theta
^{2}+\sin^{2}\theta d\phi^{2}) \, , \label{Schw}%
\end{equation}
where $\Lambda$ is the cosmological constant, $m=\frac{MG}{c^{2}}$ and $M$ is
the mass of the body. This is an exterior solution of the Einstein field
equations with cosmological constant\footnote{Throughout this work latin
indices ($a,b,...$) run from $0$ to $4$, while greek indices ($\alpha
,\beta,...$) run from $0$ to $3$.}
\begin{equation}
R_{\mu\nu}-\frac{1}{2}g_{\mu\nu}R-\Lambda g_{\mu\nu}=0 \, . \label{Einstein}%
\end{equation}
In \ the Newtonian limit of general relativity timelike geodesics of the
metric (\ref{Schw}) describes motion under a central potential
\[
V(r)=-\frac{m}{r}\text{ }-\frac{1}{6}\Lambda r^{2} \, ,
\]
with the $\Lambda$ term corresponding to a repulsive central force of
magnitude $\frac{1}{3}\Lambda r$ . According to Birkhoff's theorem
(\ref{Schw}) is the unique static solution of the Einstein field equations
(\ref{Einstein}) \cite{Rindler}.

Let us now consider the five-dimensional, static, (3-D) spherically-symmetric
metric written in coordinates ($t,r,\theta,\phi,l)$
\begin{equation}
dS^{2}=\frac{\Lambda l^{2}}{3}\left\{  \left[  1-\frac{2m}{r}-\frac{\Lambda
r^{2}}{3}\right]  dt^{2}-\left[  1-\frac{2m}{r}-\frac{\Lambda r^{2}}%
{3}\right]  ^{-1}dr^{2}-r^{2}(d\theta^{2}+\sin^{2}\theta d\phi^{2}\right\}
-dl^{2}\,, \label{Wesson}%
\end{equation}
which is a solution of the 5-D Einstein vacuum field equations
\[
R_{ab}-\frac{1}{2}g_{ab}R-\Lambda g_{ab}=0\,.
\]
We note that this is not the only possible solution for the static (3-D)
spherically-symmetric one-body problem in the presence of a cosmological
constant as Birkhoff's theorem is not valid in 5-D \cite{Wesson2}.

Now it is clear from the form of (\ref{Wesson}) that $l=\pm\sqrt{\frac
{3}{\Lambda}}$ defines two hypersurfaces which are isometric to the spacetime
(\ref{Schw}). It turns out, however, that when we take the limit
$\Lambda\rightarrow0$ in (\ref{Wesson}) the 4-D part of the metric vanishes.
As was pointed out above this may be interpreted as indicating that the
existence of a non-zero cosmological constant in $(3+1)$ may impose a rigid
relation between the $(3+1)$ and $(4+1)$-dimensional spacetimes, in turn
leading to the point of view that a cosmological constant should necessarily
exist \cite{Wesson2}). On the other hand it is not clear a priori why taking
the limit $\Lambda\rightarrow0$ $\ $in 5-D should necessarily result in such
degeneracy. In this connection it is important to recall that, contrary to the
above difficulty in 5-D, taking the limit $\Lambda\rightarrow0$ in metric
(\ref{Schw}) results in the Schwarzschild metric, which is free of such a
degeneracy. Therefore, it would be interesting if we could find a new
embedding for the static 3-D\ spherically-symmetrical one-body problem which
did not suffer from the above difficulty and for which the well-defined limit
$\Lambda\rightarrow0$ could be preserved by the process of embedding.

\section{A new embedding for the one-body metric in five dimensions}

In Ref. \cite{Wesson2}) a metric is referred to as \textit{canonical} if it is
has the form
\begin{equation}
dS^{2}=\left(  \frac{l^{2}}{L^{2}}\right)  g_{\alpha\beta}dx^{\alpha}%
dx^{\beta}-dl^{2}\,, \label{canonical}%
\end{equation}
where $g_{\alpha\beta}$ does not depend on the extra coordinate $l$ and $L$ is
a \ constant. \ The solution (\ref{Wesson}) has this form if we put
$L^{2}=\frac{3}{\Lambda}$.\ As it happens any metric of this form breaks down
when we take the limit $L\rightarrow\infty$ (which is equivalent to taking
$\Lambda\rightarrow0$ ). Therefore we should look for a vacuum solution in 5-D
that has not the canonical form in the above sense. It turns out that such a
metric can be found and is given by%

\begin{equation}
dS^{2}=l^{2}\left(  \sqrt{\frac{\Lambda w}{3}}+\frac{B}{l}\right)
^{2}\left\{  \left[  1-\frac{2m}{r}-\frac{\Lambda r^{2}}{3}\right]
dt^{2}-\left[  1-\frac{2m}{r}-\frac{\Lambda r^{2}}{3}\right]  ^{-1}%
dr^{2}-r^{2}(d\theta^{2}+\sin^{2}\theta d\phi^{2})\right\}  -wdl^{2}\,,
\label{newsolution}%
\end{equation}
where $B$ and $w$ are real constants. The solutions above describe a
Ricci-flat space and are defined for positive or negative values of $\Lambda$.
If $\Lambda$ is positive, then we take $w>0$. If $\Lambda$ is negative, we
chose $w<0$, and in this case the fifth dimension is timelike. For nonzero
$\Lambda$ it is easy to verify that for each chosen values of $w$ and $B$, the
hypersurfaces defined by
\[
l=-\sqrt{\frac{3}{w\Lambda}}(B\pm1)\,,
\]
correspond to the Schwarzschild-de Sitter spacetime (\ref{Schw})

Now taking the limit $\Lambda\rightarrow0$ \ the metric (\ref{newsolution})
becomes
\begin{equation}
dS^{2}=B^{2}\left\{  \left[  1-\frac{2m}{r}\right]  dt^{2}-\left[  1-\frac
{2m}{r}\right]  ^{-1}dr^{2}-r^{2}(d\theta^{2}+\sin^{2}\theta d\phi
^{2}\right\}  -wdl^{2}\,, \label{lambdazero}%
\end{equation}
which can be verified to be Ricci-flat. Let us denote this 5-D space by
$M_{B}^{5}$. For the particular values $B=\pm1$ and $w=1$ the equation $l=$
constant defines a trivial embedding of the 4-D Schwarzschild spacetime into
$M_{0}^{5}$ \cite{romero-etal}. Thus, the metric (\ref{newsolution}) provides
a new embedding of the 4-D Schwarzschild spacetime which is non-degenerate as
$\Lambda\rightarrow0$.

\section{Final remarks}

We recall that similar degeneracies occur, in the limit of $\Lambda
\rightarrow0$, when we consider the embedding of the de Sitter 4-D
cosmological solution
\begin{equation}
ds^{2}=dt^{2}-e^{2\sqrt{\frac{\Lambda w}{3}}t}(dx^{2}+dy^{2}+dz^{2}),
\label{deSitter}%
\end{equation}
into the 5-D space \cite{romero-etal,Wesson2,Ponce}%
\begin{equation}
dS^{2}=\frac{\Lambda}{3}l^{2}\left[  dt^{2}-e^{2\sqrt{\frac{\Lambda w}{3}}%
t}(dx^{2}+dy^{2}+dz^{2})\right]  -dl^{2}\,. \label{deSitter5}%
\end{equation}
Again it is possible to find a different 5-D Ricci-flat embedding spacetime
whose metric is not degenerate as $\Lambda\rightarrow0$. Indeed, consider the
following \ 5-D line element
\begin{equation}
dS^{2}=\left(  \sqrt{\frac{\Lambda w}{3}}l+\frac{1}{2}B\right)  ^{2}\left\{
dt^{2}-e^{2\sqrt{\frac{\Lambda w}{3}}t}(dx^{2}+dy^{2}+dz^{2})\right\}
-wdl^{2}\,, \label{newsolution2}%
\end{equation}
where, as before, $B$ and $w$ are constants\footnote{In fact both solutions
(8) and (9) correspond to 5-D flat spaces, i.e. they have vanishing Riemann
tensors.}. For nonzero $\Lambda$ one can easily verify that on the
hypersurfaces defined by
\[
l=\sqrt{\frac{3}{w\Lambda}}\left(  -\frac{B}{2}\pm1\,,\right)
\]
the metric (\ref{newsolution2}) induces the metric (\ref{deSitter}). Now
taking the limit $\Lambda\rightarrow0$ we obtain
\[
dS^{2}=\left(  \frac{1}{2}B\right)  ^{2}\left\{  dt^{2}-(dx^{2}+dy^{2}%
+dz^{2})\right\}  -wdl^{2}\,,
\]
which, depending upon whether $w=1$ or $w=-1$, represents the 5-D Minkowski
spacetime $M_{0}^{5}$ or a pseudo-Euclidean space $L^{5}$ with two timelike
dimensions. And again as we take the limit $\Lambda\rightarrow0$ the
relationship between the two manifolds is preserved in both the embedding and
the embedded manifolds. Thus taking the limit $\Lambda\rightarrow0$
(\ref{deSitter}) reduces to the 4-D Minkowski spacetime $M_{0}^{4}$, whereas
in 5-D (\ref{newsolution2}) leads to $M_{0}^{5}$ or $L^{5}$. Also clearly
$M_{0}^{4}$ is embeddable both into $M_{0}^{5}$ or $L^{5}$, with the
embeddings corresponding to $l=$ constant.

We should mention that in both cases taking $B=0$ \ and $w=1$ in
(\ref{newsolution}) and (\ref{newsolution2}) we recover the classes of
embedding spaces defined in (\ref{Wesson}) and (\ref{deSitter5}) respectively.
Regarding these spacetimes, we have found that applying the coordinate transformation%

\begin{equation}
\psi=\sqrt{w}l+B\sqrt{\frac{3}{\Lambda}}\,, \label{eq1}%
\end{equation}
transforms (\ref{canonical}) into the form (\ref{newsolution}). We note,
however, that in (\ref{eq1}) the sign of $w$ must be positive. When both
$\Lambda$ and $w$ are negative, the coordinate transformation (\ref{eq1})
becomes complex and in this case the extra coordinate $l$ is timelike and the
two metrics (\ref{canonical}) and (\ref{newsolution})\ are no longer equivalent.

Also for the metric (\ref{canonical}) the limit $\Lambda\rightarrow0$ is
degenerate whereas in (\ref{newsolution}) we have a well-defined limit. This
raises the interesting question of how a mere coordinate transformation, as in
the case of (\ref{eq1}), can so radically change the limit of a metric. This
is, in fact, a subtle and still open question and was first discussed by
Geroch \cite{Geroch}, who called attention to the fact that limits of
spacetimes may depend on the coordinate system used to perform the
calculations. In particular he showed that as the mass tends to zero the
Schwarzschild solution may have as its limits the Minkowski spacetime or
Kasner spacetime. In our present situation a simple translation in the extra
coordinate is sufficient to obtain a well behaved limit $\Lambda\rightarrow0$.

Finally, in view of the results reported here, it would be interesting to
check whether the classical tests of general relativity (in both the
Schwarzschild-de Sitter and the cosmological case) can indeed distinguish
between the conventional 4-D and the 5-D approaches if larger classes of
embedding spaces are allowed that go beyond the canonical form
(\ref{canonical}), as in the cases of (\ref{newsolution}) and
(\ref{newsolution2}) above.

\section{Acknowledgement}

We would like to thank R. Tavakol and P. S. Wesson for helpful discussions. C.
Romero thanks CNPq-FAPESQ (PRONEX) for financial support.

\end{document}